
\magnification=1200
\vsize=7.0in
\hsize=5.6in
\tolerance 10000

\baselineskip=24pt plus 2pt minus 2pt
\centerline{{\bf GAUGE INVARIANCE IN CHERN--SIMONS SYSTEMS}
 \footnote{*}{This work is supported in part by Conselho Nacional
 de Desenvolvimento Cient\'\i fico e Tecnol\'ogico, CNPq, Brazil.}}
\bigskip
\centerline{D. Bazeia}
\medskip
\centerline{Departamento de F\'\i sica, Universidade Federal
da Para\'\i ba}
\centerline{Caixa Postal 5008, 58059 Jo\~ao Pessoa, PB, Brazil}
\bigskip
\bigskip
\centerline{\bf Abstract}
\bigskip
We investigate the problem of gauge invariance of the effective
potential in Chern--Simons systems. Working at the one-loop level, we
show explicitly that the picture the subject of gauge invariance has
already constructed in scalar electrodynamics gets unchanged in the
Chern--Simons territory.
\vfill
\eject

In this paper we return$\sp{1-4}$ to the problem of gauge invariance
of the effective potential. Here the novelty appears in
three-dimensional spacetime and relies on the possibility of
introducing the Chern--Simons term. Evidently, we do not expect the
Chern--Simons term to change drastically the standard picture the
subject of gauge invariance has already constructed. Apart from its
intrinsic interest, however, a specific investigation seems
particularly desirable in view of extending former results to this
territory. To this end here we report mainly on introducing simple
and explicit arguments, which reflect the general way gauge invariance
plays it role when the standard gauge dynamics is changed to include
the Chern--Simons term.

Recent and very interesting progress on self-dual
Chern--Simons$\sp{5-8}$ and Maxwell--Chern--Simons$\sp{9,10}$ systems
have been gotten. Unfortunately, however, almost nothing has been
done concerning gauge invariance when the investigation goes beyond
the classical level.

To investigate gauge invariance of the effective potential, we then
concentrate on the Chern--Simons systems defined by
$$L\sb{CS}={1\over4}{\kappa}\,{\varepsilon\sb{\mu\nu\lambda}}
           {A\sp{\mu}}{F\sp{\nu\lambda}}+(\partial\sb{\mu}+ieA\sb{\mu})
           {\bar{\varphi}}\,(\partial\sp{\mu}-ieA\sp{\mu}){\varphi}-
           V(:\varphi:), \eqno(1)$$
and
$$L\sb{MCS}=-{1\over4}{F\sb{\mu\nu}F\sp{\mu\nu}}+L\sb{CS}. \eqno(2)$$
Here $V(:\varphi:)$ is the potential for the scalar fields, which can
be of up to sixth order in $:\varphi:$ and is supposed to present
spontaneous symmetry breaking. We notice that the above systems differ
from the ones considered in Refs.~$\lbrack{8,10}\rbrack$. Then we remark
that neither self-duality nor the presence of fermions plays any
specific role on the gauge invariance issues we shall be concerned in
the following.

To prepare the above systems to the calculation of the effective
potential, we set $\varphi = (\phi\sb1 +i\phi\sb2)/{\sqrt2}$
and use $\phi\sp2 = \phi{\sb1\sp2}+\phi{\sb2\sp2}= \phi\sb{a}\phi\sb{a}$
to write in Euclidean spacetime
$$L\sb{CS}\sp{E}=-{i\over4}{\kappa}{\varepsilon\sb{\mu\nu\lambda}}
                 {A\sb{\mu}}{F\sb{\nu\lambda}}+{1\over2}
                 {\partial\sb{\mu}}\phi\sb{a}{\partial\sb{\mu}}
                 \phi\sb{a}+{1\over2}e\sp2\phi\sp2 A\sb{\mu}A\sb{\mu}-
                 eA\sb{\mu}\varepsilon\sb{ab}\phi\sb{a}\partial\sb{\mu}
                 \phi\sb{b}+V(\phi\sp2), \eqno(3)$$
and
$$L\sb{MCS}\sp{E}={1\over4}{F\sb{\mu\nu}F\sb{\mu\nu}}+
                  L\sb{CS}\sp{E}.\eqno(4)$$
Now, under an infinitesimal gauge transformation the fields change as
$\delta A\sb{\mu}=-\partial\sb{\mu}\omega$ and
$\delta\phi\sb{a}=-e{\,}\omega{\,}\varepsilon\sb{ab}\phi\sb{b}$. Then
instead of working with (3) and (4) we have to deal with
$$L\sb{CS}\sp{eff}=L\sb{CS}\sp{E}+L\sb{g}{\,} \eqno(5)$$
and
$$L\sb{MCS}\sp{eff}={1\over4}F\sb{\mu\nu}F\sb{\mu\nu}+L\sb{CS}\sp{eff}
                    {\,}. \eqno(6)$$
Here $L\sb{g} = L\sb{f} + L\sb{c}$, and the gauge-fixing $(L\sb{f})$
and gauge-compensating $(L\sb{c})$ contributions are generically given by
$$L\sb{f}={1\over2}f\sp2(A,\phi),\phantom{DB}
  L\sb{c}=\bar{c}\,{\delta f\over{\delta\omega}}\, c. \eqno(7,8)$$
To better explore gauge invariance we choose to work with general R
gauges. In this case the gauge-fixing function is
$$f(A,\phi)={\xi\sp{-{1\over2}}}\left(\partial\sb{\mu}A\sb{\mu}+
            e{\,}\varepsilon\sb{ab}v\sb{a}\phi\sb{b}\right), \eqno(9)$$
where $\xi$ and ${\bf v}=(v\sb1,v\sb2)$ are the gauge parameters.
Then we get the gauge-fixing and gauge-compensating contributions to
$L\sb{g}$ as
$$L\sb{g}={1\over2}{\xi}\sp{-1}\left(\partial\sb{\mu}A\sb{\mu}
          +e{\,}\varepsilon\sb{ab}v\sb{a}\phi\sb{b}\right)\sp2 +
          \xi\sp{-{1\over2}}\lbrack \partial\sb{\mu}\bar{c}{\,}\partial
          \sb{\mu}c+e\sp2 v\sb{a}\phi\sb{a}\bar{c}c\rbrack. \eqno(10)$$

To calculate the effective potential one usually shifts the scalar
fields. Without loosing generality we choose to shift $\phi\sb1 \to
\bar{\phi}+ \phi\sb1$. In this case we set $v\sb1=v$ and $v\sb2=0$,
which immediately satisfies the (good gauge) condition first
introduced by Fukuda and Kugo.$\sp{11}$ The classical or zero-loop
potential is then given by $V(\bar{\phi})$. To obtain the one-loop
contributions we have to collect the quadratic terms in $L\sp{eff}$.
Here we have, after leaving out the bar over the classical field,
$$V\sb{CS}\sp{(1)}(\phi)=V\sb{H}\sp{(1)}(\phi)+V\sb{c}\sp{(1)}(\phi)+
                         V\sb{gG}\sp{(1)}(\phi), \eqno(11)$$
and
$$V\sb{MCS}\sp{(1)}(\phi)=V\sb{H}\sp{(1)}(\phi)+V\sb{c}\sp{(1)}(\phi)+
                          V\sb{MgG}\sp{(1)}(\phi), \eqno(12)$$
where $V\sb{H}\sp{(1)}(\phi)$ and $V\sb{c}\sp{(1)}(\phi)$ are the
Higgs and ghost fields contributions, respectively. They are given by
$$V\sb{H}\sp{(1)}(\phi)={1\over2}\int{d\sp3 k\over{(2\pi)\sp3}}{\,}
                        \ln\left(k\sp2 +{{d\sp2 V}\over{d\phi\sp2}}
                        \right), \eqno(13)$$
and
$$V\sb{c}\sp{(1)}(\phi)=-\int{d\sp3 k\over{(2\pi)\sp3}}{\lbrack\,
                        \ln\,(k\sp2 +e\sp2 v\phi) - {1\over2}\ln\xi
                        \,\rbrack}. \eqno(14)$$
The contributions $V\sb{gG}\sp{(1)}(\phi)$ and
$V\sb{MgG}\sp{(1)}(\phi)$ come from the gauge and Goldstone
fields, which are coupled. To get them explicitly we
write the corresponding quadratic contributions to each Lagrangian
in the general form
$${1\over2}\,\Phi\sp{t}\sb{i}M\sb{ij}\Phi\sb{j}, \eqno(15)$$
where ${\bf\Phi}\sp{t}=(A\sb1\,A\sb2\,A\sb3\,\phi\sb2)$ is the transpose
of the column vector ${\bf\Phi}$ and {\bf M} is a $4\> by\> 4$ matrix,
which can be written as, after going to momentum space,
$${\bf M}=\left(\matrix{\alpha\,\delta\sb{\mu\nu}+\beta\,
          k\sb{\mu}k\sb{\nu}+\kappa\,\varepsilon\sb{\mu\nu\lambda}
          k\sb{\lambda}&i\, e\,\xi\sp{-1}(\xi\phi - v)k\sb{\mu}\cr
          -i\, e\,\xi\sp{-1}(\xi\phi - v)k\sb{\nu}&
          k\sp2 + (1/{\phi})(dV/{d\phi})\cr}\right). \eqno(16)$$
Here we recall that in the 't Hooft/R$\sb{\xi}$ gauge $v=\xi\,\phi$.
In this case the gauge-Goldstone coupling vanishes, as we can
immediately see from $(16)$. In the general case wethen recognize that
the Goldstone term, $k\sp2 + (1/{\phi})(dV/{d\phi})$, does not depend on
the particular Chern--Simons or Maxwell--Chern--Simons system one is
considering. This is also true for
$i\, e\,\xi\sp{-1}(\xi\phi - v)\, k\sb{\mu}$, the coupling between the
gauge and Goldstone fields. The gauge term has the general structure
$\alpha\,\delta\sb{\mu\nu}+\beta\,k\sb{\mu}k\sb{\nu}+
\kappa\,\varepsilon\sb{\mu\nu\lambda}\, k\sb{\lambda}$. However,
$\alpha$ and $\beta$ depend on the particular model one is working with:
for the Chern--Simons system we have
$$\alpha\sb{CS}=e\sp2 \phi\sp2,\phantom{DB}
   \beta\sb{CS}=\xi\sp{-1}; \eqno(17a,b)$$
for the Maxwell--Chern--Simons system they are given by
$$\alpha\sb{MCS}= k\sp2 + e\sp2 \phi\sp2,\phantom{DB}
   \beta\sb{MCS}=\xi\sp{-1}(1 - \xi). \eqno(18a,b)$$
Now, after calculating the determinants we get
$$V\sb{gG}\sp{(1)}={1\over2}\int{d\sp3 k\over{(2\pi)\sp3}}\lbrace\,\ln(
                   \kappa\sp2 k\sp2+e\sp4\phi\sp4)+\ln\lbrack\,(k\sp2+
                   e\sp2 v\,\phi)\sp2+{1\over{\phi}}{dV\over{d\phi}}(
                   k\sp2+\xi e\sp2\phi\sp2\,)\,\rbrack -\ln\xi\,\rbrace,
                   \eqno(19)$$
and
$$V\sb{MgG}\sp{(1)}={1\over2}\int{d\sp3 k\over{(2\pi)\sp3}}\lbrace\,\ln
                    \lbrack\,\kappa\sp2 k\sp2+(k\sp2+e\sp2\phi\sp2)
                    \sp2\,\rbrack+\ln\lbrack\,(k\sp2+e\sp2 v\,\phi)\sp2+
                    {1\over{\phi}}{dV\over{d\phi}}(k\sp2+\xi e\sp2\phi
                    \sp2\,)\,\rbrack -\ln\xi\,\rbrace. \eqno(20)$$

To write the effective potentials to the Chern--Simons and
Maxwell--Chern--Simons systems we collect the results already obtained
to get, up to the one-loop order
$$\eqalign{V\sb{CS}
      \sp1 (\phi)&=V(\phi)+{1\over2}\int{d\sp3 k\over{(2\pi)\sp3}}
                   \lbrace\,\ln\left(\, k\sp2 +{{d\sp2 V}
                   \over{d\phi\sp2}}\right)+\ln\left( k\sp2 +
                   {{e\sp4\phi\sp4}/{\kappa\sp2}}\right)+\cr
                   &\qquad\ln\lbrack( k\sp2+e\sp2 v\,\phi\, )\sp2+
                   {1\over{\phi}}{dV\over{d\phi}}( k\sp2+
                   \xi e\sp2\phi\sp2\,)\rbrack-\ln( k\sp2+
                   e\sp2 v\,\phi\, )\sp2\, \rbrace, \cr} \eqno(21)$$
and
$$\eqalign{V\sb{MCS}
        \sp1 (\phi)&=V(\phi)+{1\over2}\int{d\sp3 k\over{(2\pi)\sp3}}
                     \lbrace\,\ln\left( k\sp2 +{{d\sp2 V}
                     \over{d\phi\sp2}}\right)+\ln( k\sp2+m\sb{+}\sp2)+
                     \ln(k\sp2+m\sb{-}\sp2)+\cr
                     &\qquad\ln\lbrack{\,}
                     ( k\sp2+e\sp2 v\,\phi\, )\sp2+{1\over{\phi}}
                     {dV\over{d\phi}}( k\sp2+\xi e\sp2\phi\sp2\,)\,
                     \rbrack-\ln( k\sp2+e\sp2 v\,\phi\, )\sp2
                     \, \rbrace. \cr} \eqno(22)$$
In the above result we have set $\kappa\sp2 k\sp2+( k\sp2+
e\sp2\phi\sp2\, )\sp2 = (k\sp2+m\sb{+}\sp2)( k\sp2+m\sb{-}\sp2)$; then
$$m\sb{\pm}\sp2=e\sp2\phi\sp2 +{1\over2}\kappa\sp2 \pm {1\over2}\kappa
                \sp2\sqrt{1 + 4 e\sp2 \phi\sp2/\kappa\sp2}. \eqno(23)$$

To investigate gauge invariance we now use $(21)$ and $(22)$. Here we
immediately see that both results present the same gauge-dependent
contribution; explicitly
$$V\sp{(1)}(\phi;\xi,v)={1\over2}\int{d\sp3 k\over{(2\pi)\sp3}}
                        \lbrace\,\ln\lbrack\, ( k\sp2+e\sp2 v\,\phi)
                        \sp2+{1\over{\phi}}{dV\over{d\phi}}( k\sp2+
                        \xi e\sp2\phi\sp2\,)\,\rbrack-\ln( k\sp2+
                        e\sp2 v\,\phi)\sp2\, \rbrace. \eqno(24)$$
More importantly, this result is exactly what we have already
found$\sp3$ in standard scalar electrodynamics. Then the proof
of gauge invariance of the effective potential is already given in
Refs~$\lbrack{3,4}\rbrack$. Here we recall that in the
Nielsen$\sp{1}$ way to check gauge invariance the quantities$\sp{3,4}$
$$C\sb{\xi}(\phi;\xi,v)=\xi\,{\partial\phi\over{\partial\xi}},
   \phantom{DB}
    C\sb{v}(\phi;\xi,v)=v\,{\partial\phi\over{\partial v}},\eqno(25a,b)$$
play a basic role in constructing the identities
$$\xi\,{dV\over{d\xi}}=\xi\,{\partial V\over{\partial\xi}}+C\sb{\xi}\,
                       {\partial V\over{\partial\phi}}=0 , \eqno(26a)$$
and
$$v\,{dV\over{dv}}=v\,{\partial V\over{\partial v}}+C\sb{v}\,
                   {\partial V\over{\partial\phi}}=0 , \eqno(26b)$$
which ensure gauge invariance of the effective potential. This subject
has been extensively discussed in the past.$\sp{1-4}$ Then we omit
details to write, up to the one-loop order,
$$C\sb{\xi}\sp{(1)}
       (\phi;\xi,v)=-{1\over2}\xi e\sp2\phi\int{d\sp3 k\over{(2\pi)
                    \sp3}}\lbrace\, 1\,/\,\lbrack\,(k\sp2+e\sp2 v\phi)
                    \sp2+{1\over\phi}{dV\over{d\phi}}(k\sp2 +\xi e\sp2
                    \phi\sp2)\,\rbrack\,\rbrace\eqno(27a)$$
and
$$C\sb{v}\sp{(1)}
     (\phi;\xi,v)=e\sp2 v\int{d\sp3 k\over{(2\pi)\sp3}}\lbrace(k\sp2+
                  \xi e\sp2\phi\sp2)\,/\,(k\sp2+e\sp2 v\phi)\lbrack(k
                  \sp2+e\sp2 v\phi)\sp2+{1\over\phi}{dV\over{d\phi}}(k
                  \sp2+\xi e\sp2\phi\sp2)\rbrack\,\rbrace\eqno(27b)$$
Note that both $C\sb{\xi}\sp{(1)}(\phi;\xi,v)$ and
$C\sb{v}\sp{(1)}(\phi;\xi,v)$ are finite in three dimensions, although
renormalization does not change our conclusions.$\sp{3,4}$

To understanding why gauge invariance goes the above way, we recall that
the BRST symmetry the effective Lagrangian $(6)$ engenders remains
unchanged when one discards the Maxwell or the Chern--Simons term.
However, in order not to change the way we are doing this investigation,
we simply note that $C\sb{\xi}\sp{(1)}(\phi;\xi,v)$ and
$C\sb{v}\sp{(1)}(\phi;\xi,v)$ only depend on the ghost, Goldstone and
gauge-Goldstone propagators.$\sp{3,4}$ And these propagators do not
change when one goes from standard scalar electrodynamics to
Chern--Simons scalar electrodynamics. To make this point clear, let us
now write down the propagators. Here after some algebraic manipulations
we get, for the Higgs field
$$\Delta\sb{H}( k)={1\over{k\sp2+d\sp2 V/d\phi\sp2}}, \eqno(28)$$
for the ghost field
$$\Delta\sb{g}( k)=\xi\sp{-{1\over2}}D\sb{v}\sp{-1}( k), \eqno(29)$$
for the Goldstone field
$$\Delta\sb{G}( k)=D\sb{v\xi}\sp{-1}( k)\,D\sb{\xi}( k), \eqno(30)$$
for the gauge-Goldstone field
$$\Delta\sb{G\mu}( k)=i\, e\,(\xi\phi-v)\, D\sb{v\xi}\sp{-1}
                      ( k)\, k\sb{\mu}, \eqno(31)$$
for the gauge field
$$\eqalign{\Delta\sb{\mu\nu}
       \sp{CS}( k)&={e\sp2\phi\sp2\over{\kappa\sp2}}D\sb{\kappa}\sp{-1}
                    ( k)\lbrace\,\delta\sb{\mu\nu}+
                    (\xi-1)D\sb{\xi}\sp{-1}( k)\, k\sb{\mu}k\sb{\nu}
                    -{\kappa\over{e\sp2\phi\sp2}}\varepsilon
                    \sb{\mu\nu\lambda}k\sb{\lambda}+\cr
                    &\qquad\xi\left({{\kappa
                    \sp2}\over{e\sp2\phi\sp2}}-1\right)D\sb{v\xi}
                    \sp{-1}( k)D\sb{\xi v}( k)k\sb{\mu}
                    k\sb{\nu}+e\sp2(\xi\phi-v)\sp2 D\sb{v\xi}\sp{-1}
                    ( k)D\sb{\xi}\sp{-1}( k)D( k)
                    k\sb{\mu}k\sb{\nu}\rbrace, \cr} \eqno(32)$$
and
$$\eqalign{\Delta\sb{\mu\nu}
  \sp{MCS}( k)&=D\sb{+}\sp{-1}( k)D\sb{-}\sp{-1}( k)
                D( k)\,\lbrace\,\delta\sb{\mu\nu}+(\xi-1)
                D\sb{\xi}\sp{-1}( k)k\sb{\mu}k\sb{\nu}-\kappa
                D\sp{-1}( k)\varepsilon\sb{\mu\nu\lambda}k
                \sb{\lambda}+\cr
                &\qquad \xi\kappa\sp2 D\sb{v\xi}\sp{-1}( k)
                D\sp{-1}( k)D\sb{\xi v}( k)k\sb{\mu}k\sb{\nu}+
                e\sp2(\xi\phi-v)\sp2 D\sb{v\xi}\sp{-1}( k)
                D\sb{\xi}\sp{-1}( k) D( k)k\sb{\mu}k
                \sb{\nu}\,\rbrace. \cr} \eqno(33)$$
For simplicity, in the above results we have set
$$D( k)=k\sp2+e\sp2\phi\sp2,\phantom{DB}
  D\sb{\kappa}( k)=k\sp2+e\sp4\phi\sp4/\kappa\sp2, \eqno(34a,b)$$
$$D\sb{\xi}( k)=k\sp2+\xi e\sp2\phi\sp2,\phantom{DB}
  D\sb{v}( k)=k\sp2+e\sp2 v \phi, \eqno(34c,d)$$
$$D\sb{+}( k)=k\sp2+ m\sb{+}\sp2,\phantom{DB}
  D\sb{-}( k)=k\sp2+ m\sb{-}\sp2, \eqno(34e,f)$$
$$D\sb{v\xi}( k)=( k\sp2+e\sp2 v \phi)\sp2 + {1\over{\phi}}
                 {dV\over{d\phi}}(k\sp2+\xi e\sp2\phi\sp2),\eqno(34g)$$
and
$$D\sb{\xi v}( k)=k\sp2+\xi\sp{-1} e\sp2 v\sp2+
                  {1\over{\phi}}{dV\over{d\phi}}. \eqno(34h)$$
Then we note that the propagator for the gauge field is the only to
depend on the particular system we are working with; the others do not
change anything and, more importantly, they are equal to the
corresponding propagators one gets in standard scalar
electrodynamics$\sp3$.

Some remarks are now in order: First, in $\Delta\sb{\mu\nu}\sp{MCS}$
put $\kappa = 0$ to reproduce the result one has in standard scalar
electrodynamics.$\sp3$ Second, yet in $\Delta\sb{\mu\nu}\sp{MCS}$
use $v = \xi\phi$ to obtain the result in the 't Hooft/R$\sb{\xi}$
gauge.$\sp{12}$. Third, set $v = 0$ and $\xi = 0$ in
$\Delta\sb{\mu\nu}\sp{CS}$ to get the result in Landau gauge;
here, compare with Ref.~\lbrack 8\rbrack, apart from the difference
in the spacetime metric and in the way the scalar field is there
shifted.

To conclude, we have explicitly shown that the gauge-dependent
contribution to the effective potential at the one-loop level remains
unchanged when one goes from standard scalar electrodynamics to
Chern--Simons scalar electrodynamics, despite the presence of the
Maxwell term in the last case. Then the proof of gauge invariance
of the effective potential follows in the same way it does when
the gauge field presents standard dynamics.

As a final comment, we recall that the Chern--Simons term gets a
surface contribution after a gauge change. Then we have to care about
the boundary conditions at the border of the underlying manifold one
is working with. Such a problem appears, for instance, when thermal
effects are introduced, but this was already considered in
Ref.~\lbrack13\rbrack.

\vfill
\eject
\centerline{\bf REFERENCES}
\medskip
\item{1.}N. K. Nielsen, Nucl. Phys. {\bf B101} (1975) 173.
\medskip
\item{2.}I. J. R. Aitchison and C. M. Fraser, Ann. Phys. (NY) {\bf 176}
(1984) 1.
\medskip
\item{3.}J. R. S. do Nascimento and D. Bazeia, Phys. Rev. {\bf D35}
(1987) 2490.
\medskip
\item{4.}A. F. de Lima and D. Bazeia, Z. Phys. {\bf C45} (1989) 471.
\medskip
\item{5.}J. Hong, Y. Kim and P. Y. Pac, Phys. Rev. Lett.{\bf 64}
(1990) 2230.
\medskip
\item{6.}R. Jackiw and E. J. Weinberg, Phys. Rev. Lett. {\bf 64}
(1990) 2234.
\medskip
\item{7.}R. Jackiw, K. Lee and E. J. Weinberg, Phys. Rev. {\bf D42}
(1990) 3488.
\medskip
\item{8.}Y. Ipeko\v glu, M. Leblanc and M. T. Thomaz, Ann. Phys. (NY)
{\bf 214} (1992) 160.
\medskip
\item{9.}C. Lee, K. Lee and H. Min, Phys. Lett. {\bf B252} (1990) 79.
\medskip
\item{10.}M. Leblanc and M. T. Thomaz, Phys. Rev. {\bf D}, to appear,
and MIT report, CTP \# 2042, (January, 1992).
\medskip
\item{11.}R. Fukuda and T. Kugo, Phys. Rev. {\bf D13} (1976) 3469.
\medskip
\item{12.}R. D. Pisarski and S. Rao, Phys. Rev. {\bf D32} (1985) 2081.
\medskip
\item{13.}M. T. Thomaz, MIT report, CTP \# 2119 (July, 1992).
\vfill
\eject
\bye